\documentclass[lettersize,journal]{IEEEtran}
\pdfoutput=1
\usepackage{amsmath,amsfonts}
\usepackage{algorithmic}
\usepackage{algorithm}
\usepackage{array}
\usepackage[caption=false,font=normalsize,labelfont=sf,textfont=sf]{subfig}
\usepackage{textcomp}
\usepackage{stfloats}
\usepackage{url}
\usepackage{verbatim}
\usepackage{graphicx}
\usepackage{cite}
\usepackage{color}
\begin{document}
\title{A high-sensitivity frequency counter for free-induction-decay signals}

\author{

Tong Gong,
\thanks{Tong Gong ,Jiang He and Dong Sheng are with Department of Precision Machinery and Precision Instrumentation, Key Laboratory of Precision Scientific Instrumentation of Anhui Higher Education Institutes, University of Science and Technology of China, Hefei 230027, China}
Ming-Rui Shu,
\thanks{Ming-Rui Shu is with China Ship Research and Design Center, Wuhan 430000, China}
Jiang He,
Kai Liu,
Yi-Ren Li,
Xin-Jun Hao,
Dong Sheng,
Yu-Ming Wang,
\thanks{Kai Liu, Yi-Ren Li, Xin-Jun Hao and Yu-Ming Wang are with the School of Earth and Space Sciences, and Deep Space Exploration Laboratory, University of Science and Technology of China \& CAS Center for Excellence in Comparative Planetology, and CAS Key Laboratory of Geospace Environment, Hefei 230026, China. (Yu-Ming Wang's email: ymwang@ustc.edu.cn)}
Yu-Kun Feng
\thanks{Yu-Kun Feng is with the Department of Modern Physics, University of Science and Technology of China, Hefei 230026, China. (Yu-Kun Feng's email: fyk@ustc.edu.cn)}
}

\maketitle

\begin{abstract}

Real-time frequency readout of time-dependent pulsed signals with a high sensitivity are key elements in many applications using atomic devices, such as FID atomic magnetometers. In this paper, we propose a frequency measurement algorithm based on the Hilbert transform and implement such a scheme in a FPGA-based frequency counter. By testing pulsed exponential-decay oscillation signals in the frequency range of 10 to 500 kHz, this frequency counter shows a frequency sensitivity better than 100 $ \mathbf{\mu Hz/\sqrt {Hz}} $ at 10 Hz, with an output rate of 200 Hz. When the output rate is increased to 1000 Hz, the sensitivity remains better than 400 $ \mathbf{\mu Hz/\sqrt {Hz}} $ at 10 Hz. The performance on frequency sensitivity is comparable with results obtained by off-line nonlinear fitting processes. In addition, this frequency counter does not require the pre-knowledge of the analytic expression of the input signals. The realization of such a device paves the way for practical applications of highly-sensitive FID atomic magnetometers.
  
\end{abstract}

\begin{IEEEkeywords}
Frequency counter, Hilbert transform, High-sensitivity, FID magnetometer, High bandwidth.
\end{IEEEkeywords}

\section{Introduction}

Atomic devices use atomic transitions for measurements of observables, and an important application of atomic devices is to measure magnetic fields \cite{5778937,kitching2018chip}. In recent years, atomic magnetometers have undergone rapid developments, with the sensitivity for geomagnetic measurements surpassing traditional fluxgate magnetometers \cite{budker2007optica}. The principle of these magnetometers involves converting magnetic field measurements into frequency measurements by utilizing the Larmor precession of polarized atoms in an external magnetic field \cite{grujic2015sensitive}. For example, in a rubidium atomic magnetometer, the dynamic range within the geomagnetic field is approximately 65000 nT, with corresponding output frequencies less than 500 kHz. Therefore, the resolution of frequency measurement of the sensor directly affects the precision of magnetic field measurement. For atomic magnetometers working in the closed-loop mode \cite{abel2020optically}, the signals are continuous oscillations with constant amplitudes. To achieve better accuracy, a pulsed working mode, which is often referred as the free-induction-decay (FID) mode, is developed, where the amplitude of the signals are time-varying. For most geophysics applications, the detection bandwidth of the magnetometer is normally required to be as large as possible. Therefore, it is necessary to develop a high-precision and high-bandwidth frequency measurement method.

In previous researches with FID magnetometers \cite{PhysRevLett.110.160802}, the most reliable method of data analysis involves fitting the signal with a known function to obtain its frequency parameters. However, nonlinear fittings are computationally expensive. For instance, applying the Levenberg-Marquardt (LM) \cite{more2006levenberg} algorithm to a segment of damped oscillating signal data with a size of 1000 points on a personal computer with a CPU of i7-6700 may requires over 0.1 second for fitting. Considering computational speed and hardware limitations for miniaturized devices, it is desired to find frequency processing algorithms with lower computational loads. Another commonly used method relies on pulse-counting method, including a measuring-cycle method in relatively low-frequency bands and a measuring-frequency method in relatively high-frequency bands \cite{keshani2020digital}. However, such methods have intrinsic problems as quantization errors. This quantization error is practically limited by the hardware (the clock frequency and sampling rate) and the sensor bandwidth (measurement time). To address this problem, various optimization schemes have been proposed. For example, the equal-precision method reduces quantization errors during signal counting \cite{kumar2011new}, while the multi-channel method uses multiple counting channels to average quantization errors and enhance the measurement accuracy \cite{dong2016high}. The multi-phase clock method suppresses quantization errors by utilizing multiple reference clocks with different phases simultaneously \cite{ge2024high}. The relative-frequency-difference method employs high-precision time-to-digital converters (TDC) to directly measure tiny time intervals to reduce the quantization errors \cite{samarah2013digital}, and the $\Omega$-counter represents the state-of-art optimization of this approach, leveraging continuous timestamp registration and linear regression algorithms to extract frequency parameters with minimal computational bias \cite{kwiatkowski2024hardware}. However, these optimization methods either marginally improve measurement accuracy or consume excessive hardware resources. In addition, another widely used fast frequency counting method is based on spectrum analysis, such as the discrete Fourier transform \cite{visschers2021rapid}. The accuracy of spectral analysis depends on the sampling rate and data volume, it has been found in practice that the sensitivity based on this method is two times worse than that using the nonlinear fitting method \cite{PhysRevApplied.22.014084}.

In this paper, we present a frequency measurement algorithm based on the Hilbert transform. This algorithm converts nonlinear frequency fitting to linear phase fitting without sacrificing computational accuracy, therefore simultaneously satisfies the requirements on extracting the signal oscillation frequency with a high precision and bandwidth. Schemes based on similar algorithms have been demonstrated to work in principle \cite{PhysRevResearch.2.013213,hunter2022accurate}. In this work, we show that the algorithm needs to be modified for fast and precise real-time analysis applications. Furthermore, we develop a hardware system to implement this algorithm, and test its performance using pulsed signals from arbitrary function generators. The algorithm and hardware system developed in this work is not only important for atomic devices, but also can be applied in any field that requires a reliable real-time frequency extraction from pulsed oscillation signals, such as underwater acoustic signal processing \cite{stojanovic2009underwater}. The structure of the remainder of this paper is as follows: Section II provides working principle of the algorithm and the hardware design, Section III presents the test results of this algorithm and hardware, and Section IV concludes this paper.

\section{Algorithm and hardware design}

The Hilbert transform is widely used in signal processing, with a physical interpretation of delaying the signal by $ 90^\circ $ in all frequency components. For a continuous-time signal $ x(t) $, its Hilbert transform is denoted as $ y(t)=\frac{1}{\pi } \int_{- \infty }^{\infty } \frac{x(\tau )}{t-\tau } \mathrm{d}\tau $. If the original signal can be represented as $ x(t)=\int_{0}^{\infty } A(\omega )\cos (\omega t+\varphi (\omega ))\mathrm{d}\omega $, then its Hilbert transform results in $ y(t)=\int_{0}^{\infty } A(\omega )\sin (\omega t+\varphi (\omega ))\mathrm{d}\omega $. By using the original signal as the real part and its Hilbert transform as the imaginary part, a complex analytic signal $z(t) = x(t) + iy(t)$ can be constructed. For the signal $ z(t) $, its instantaneous amplitude $ A(t)=\sqrt{x^2(t)+y^2(t)} $ and instantaneous phase $ \varphi (t)=\tan^{-1} (y(t)/x(t)) $ at any given moment can be constructed, where the derivative of the instantaneous phase with respect to time yields the instantaneous frequency. In practice, signals are discrete, and the corresponding discrete-time Hilbert transform is represented as \cite{2006Signals}

\begin{equation}
    \label{1}
    y[n]=\frac{2}{\pi} \sum_{}^{}  \frac{x[n-k]}{k}\sin ^2\left (\frac{k \pi}{2}\right )  , k \in Z
\end{equation}
 
Suppose the discrete signal takes the form $ x[n]=A[n]\sin [\omega n/f_s] $, where $ f_s $ is the sampling rate, the corresponding $ y[n] $ can be expressed as

\begin{equation}
    \label{2}
    \begin{aligned}
        y[n]&=\frac{2}{\pi} \sum_{}^{}  \frac{A[n-k]\sin [\omega (n-k)/f_s]}{k} \\
        &\quad-\frac{2}{\pi} \sum_{}^{}  \frac{A[n+k]\sin [\omega (n+k)/f_s]}{k} \\
        &=\frac{2}{\pi} \sum_{}^{}  \frac{(A[n+k]-A[n-k])\sin [\omega (n+k)/f_s]}{k} \\
        &\quad-\frac{2}{\pi} \sum_{}^{}  \frac{2A[n-k]\cos [\omega n/f_s]\sin [\omega k/f_s]}{k}, \\ 
        &\qquad k=1,3,5,\dots
    \end{aligned}
\end{equation}

Considering the practical range of signal amplitudes where $ A[n] \le  A_{max} $, the expression for $ y[n] $ converges for each $ n $. Therefore, when calculating the discrete-time Hilbert transform, it is feasible to truncate at a suitable value of k to reduce computational complexity. The algorithm developed in this work, which is named as Hilbert-Transform Linear-Regression (HT-LR) algorithm, is outlined below:

\begin{itemize}
    \item[(1)]
    Utilize Eq.(1) to transform the acquired signal $ x[n] $ into $ y[n] $.
    \item[(2)]
    Calculate the instantaneous phase $ \varphi [n]=\tan^{-1} (y[n]/x[n]) $ and derive the cumulative phase $ \Phi [n]=\varphi [n]+2 \pi N $ to make phase value continuous.
    \item[(3)]  
    Perform a weighted linear regression fit on $ \Phi (t) $ with the amplitude $ z[n]=\sqrt{x[n]^{2}+y[n]^{2}} $ as the weight where $ t=n/f_s $, and the extracted slope represents the oscillation frequency of the signal.
\end{itemize}

\begin{figure}[htbp]
    \centering
    \includegraphics[width=0.48\textwidth]{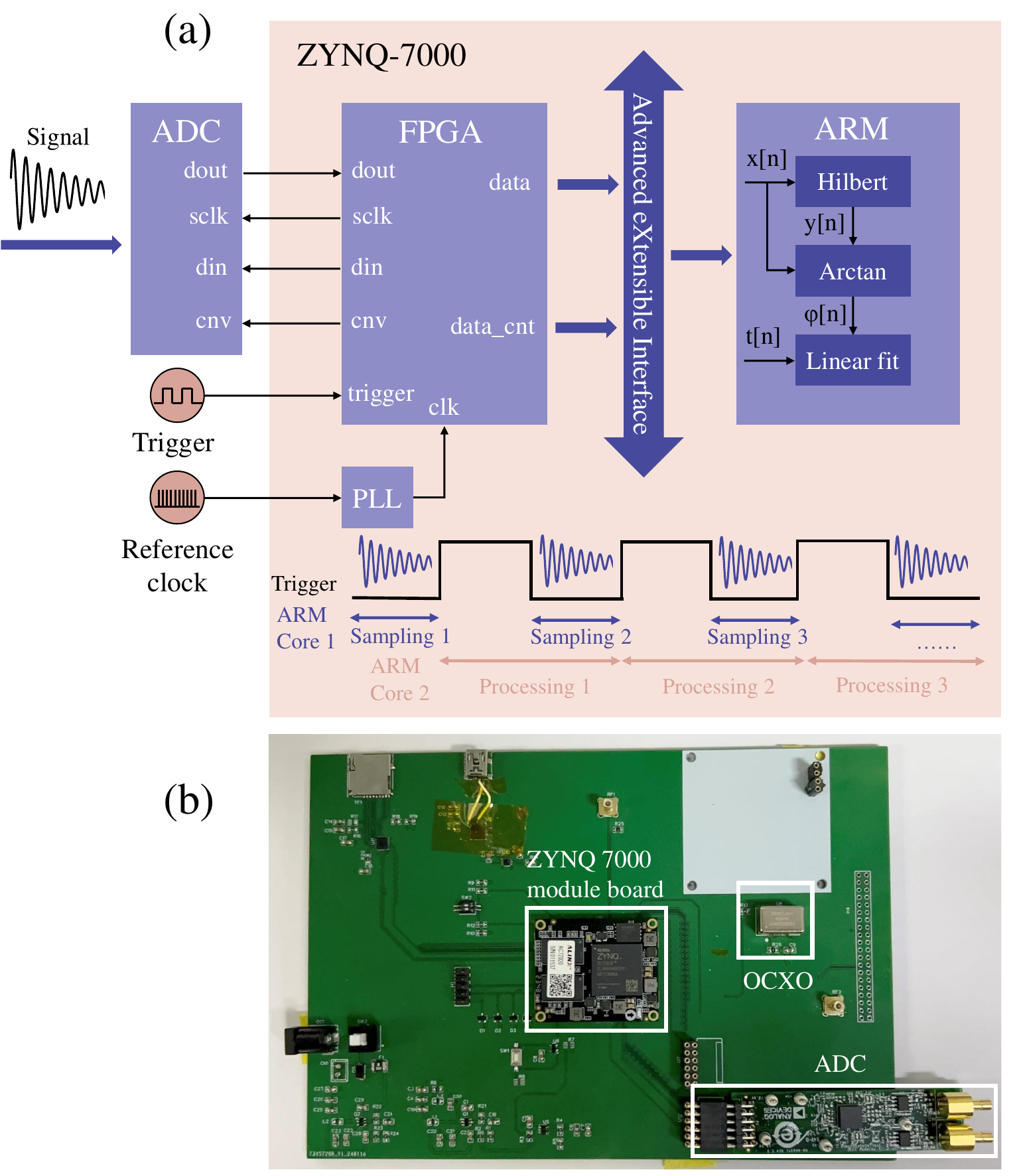}
    \caption{(a) Schematic diagram of the frequency counter. (b) A picture of the hardware. The circuit board includes an additional ADC interface reserved for the differential configuration of future atomic magnetometers, which was not utilized in the tests described in this work.}
    \label{fig1}
\end{figure}

The HT-LR algorithm is implemented in a Field-Prgrammable Gate Array (FPGA) based frequency counter. The hardware of this frequency counter consists of four main parts:  analog-to-digital converter module (ADC), communication interface, a frequency reference, and the main control chip, as shown in Fig. \ref{fig1}. Additionally, we use a 16-bit signal generator to generate signals to test the performance of the frequency counter. In this hardware, we use 18-bit ADCs with a maximum sampling rate of 2 MSa/s. The frequency reference is an Oven-Controlled Crystal Oscillator (OCXO, 40 MHz, 50 ppb accuracy), which provides the reference clock for the frequency counter. The main control chip (ZYNQ-XC7Z020) consists of two parts: the Processing System (PS) and the Programmable Logic (PL). The signal acquisition module is implemented on the PL side, controlling ADC signal acquisition of the test signal using a reference clock and trigger signal. The data processing module is implemented on the PS side, processing the acquired data, and mainly consists of three calculation units: Discrete Hilbert Transform, arctangent operation for cumulative phase calculation, and least squares linear fitting. The ZYNQ chip's PS system integrates two ARM Cortex-A9 processors, and each core can independently run different tasks. At the beginning of each acquisition processing cycle, one processor receives data from the PL side and stores it in the ZYNQ's On-Chip Memory (OCM). Once all data is stored, another processor retrieves the data from OCM for processing.

The hardware architecture was designed with the following considerations: 

1. The ZYNQ-XC7Z020 was chosen over conventional microprocessors such as STM32 due to its hybrid architecture combining PL and dual ARM Cortex-A9 processors (PS), while general-purpose microprocessors struggle to meet real-time processing requirements for Hilbert transforms at high sampling rate.

2. Standalone 18-bit ADCs were adopted because their simultaneous low noise and 2 MSa/s sampling rate exceed the performance of embedded ADC blocks in typical FPGA platforms.

3. Partitioning data acquisition (ARM Core 1) and signal processing (ARM Core 2) across two CPU cores provides 50\% temporal margin for computational tasks.

4. The integrated ARM cores provide a mature software ecosystem for algorithm implementation, while the FPGA enables hardware acceleration. This PS-PL co-processing framework allows seamless software-hardware co-design: complex signal processing chains can be initially prototyped on the ARM cores and later gradual migration to programmable logic.

\section{Test Results}

The simulated signals take the form of damped oscillating pulses $ Ae^{-t/\tau}\sin (2\pi ft) $, where signal amplitude $ A = 2.5 $ V, relaxation time $ \tau = 2.5 $ ms, $ f $ is in the range of 10 kHz to 500 kHz, and the repetition rate of the signal is 200 Hz with a duty cycle of 50\%. The upper frequency limit of the test signals was constrained by the ADC's maximum sampling rate, while this frequency range also corresponds to the operational bandwidth of atomic magnetometers within geomagnetic field environments. These test signals were generated by a RIGOL DG2052 function generator, which simultaneously provided a synchronized 200 Hz square-wave trigger signal. The trigger signal synchronizes ADC acquisition with the pulsed FID signal. For the application of magnetometers, it ensuring alignment between data segments and the probe/pumping cycle of the magnetometer. To be specific, the trigger signal was held at low-level during the output of damped oscillation pulses and switched to high-level during the quiescent periods, thereby controlling the timing sequence in our frequency counter. The ADC sampling rate of the frequency counter is set to 1.53846 MSa/s (sampling interval of 650 ns). While processing data using the HT-LR algorithm, the ADC module simultaneously store the acquired data on a memory card so that off-line nonlinear fitting of the same data can be performed for comparisons.

In practice, the infinite series sum in the discrete Hilbert Transform (Eq. \ref{2}) needs to be truncated. This leads to the sum over $ k $ in Eq. (\ref{1}) with a maximum value of $ K $ 

\begin{equation}
    \label{3}
    y[n]=\frac{2}{\pi} \sum_{-K }^{K }  \frac{x[n-k]}{k} , k=\pm1, \pm3, \pm5, \cdots
\end{equation}

The precision of extracted frequency increases with the value of $ K $, but this also increases computation time. In this work, we use the noise spectral density (NSD) value at 10 Hz (averaged between 8-12 Hz) of the extracted frequency as a main parameter to characterize the performance of the frequency counter. 

\begin{figure}[htbp]
    \centering
    \includegraphics[width=0.48\textwidth]{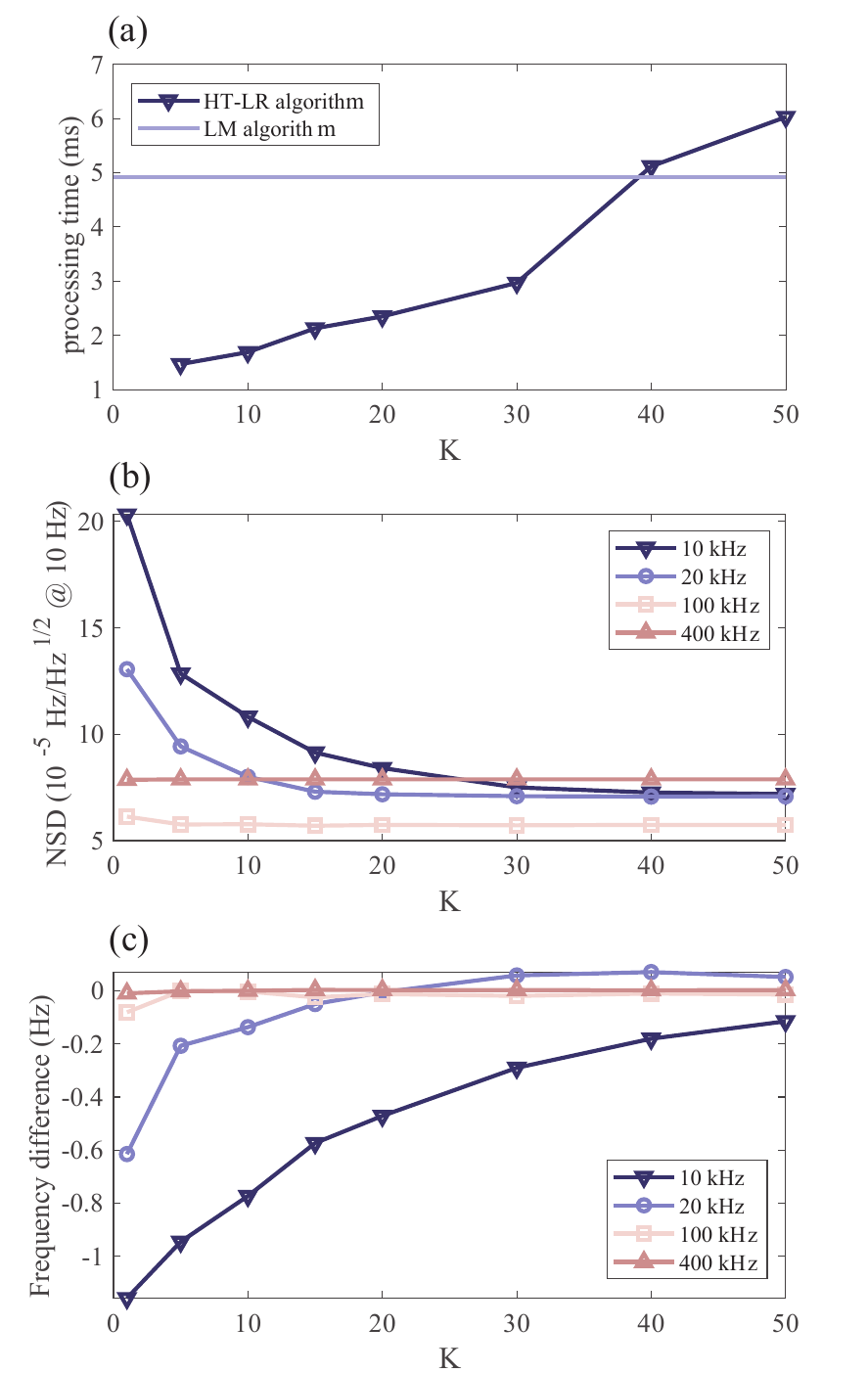}
    \caption{(a) Computation time of the HT-LR algorithm for different $ K $ values, with the horizontal line refers to the time required for the same data processed offline by nonlinear fitting using MATLAB on a computer with a CPU of i9-139000. To benchmark efficiency, we implemented the HT-LR algorithm in MATLAB and observed its execution time on the same workstation (i9-13900 CPU) to be merely 1/20 of that on the FPGA platform. (b) NSD value (at 10 Hz) of the frequency counter output for different $ K $ values. (c) Difference $ \Delta f $ between the frequency extracted with different $ K $ values with the value of extracted by off-line fitting using the LM algorithm.}
    \label{fig2}
\end{figure}

Figure \ref{fig2}(a) shows that the computation time for pulsed signals with a duration of 2.5 ms changes with different values of $ K $. A horizontal line at 4.92 ms represents the time required for the same data processed by running the LM fitting function in MATLAB on a computer with a CPU of i9-13900. It can be observed that when $ K $ is less than 40, the signal processing speed of our frequency counter exceeds that of the nonlinear fitting algorithm running on a high-performance computer. Considering the hardware limitations and performance optimizations, performing nonlinear fitting on home-made data acquisition systems would require even longer time.

To systematically characterize the measurement performance, we experimentally quantified the NSD and $ \Delta f $ of our frequency counter. While the mathematical theory of Hilbert-transform-based methods for FID signal processing has been rigorously established in prior works \cite{arXiv}, this study focuses on error analysis under practical operating conditions. As shown in Fig. \ref{fig2}(b) and \ref{fig2}(c), as $ K $ increases, the sensitivity of the HT-LR algorithm gradually converges, so does the difference between the extracted frequencies by the HT-LR and LM algorithms. Notably, the $ \Delta f $ induced by the truncation parameter $ K $ constitutes the primary source of systematic error in this algorithm, though such error is one order of magnitude smaller than the typical intrinsic heading errors of the FID magnetometers used in geomagnetic fields \cite{PhysRevA.111.023119}. We choose $ K=20 $ in the rest of the tests, which strikes a balance between computational efficiency and precision.

\begin{figure}[htbp]
    \centering
    \includegraphics[width=0.48\textwidth]{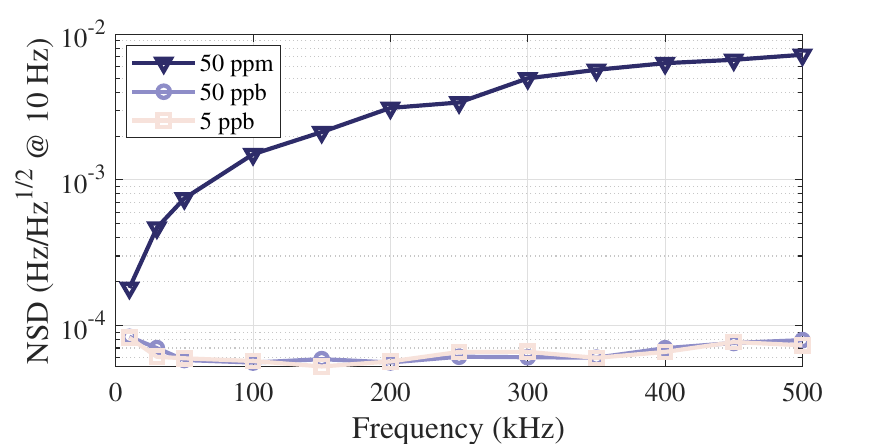}
    \caption{Comparison of frequency counter results at different signal frequencies using oscillators with varying stability, with $ K=20 $ in the HT-LR algorithm.}
    \label{fig3}
\end{figure}

The stability of crystal oscillators affects the sensitivity of frequency measurements because it determins the stability of the sampling frequency $ f_s $. We compare the performance of the frequency counter when the time reference of the FPGA is its onboard crystal oscillator (50 ppm accuracy) with the cases where the frequency references are based on two different OCXOs (50 ppb and 5 ppb accuracy, respectively), as shown in Fig. \ref{fig3}. It can be concluded that for the tested signal frequency range (10–500 kHz) and relaxation time ($\tau = 2.5$ ms), once the stability of the crystal oscillator exceeds 50 ppb, the sensitivity of the frequency counter is no longer limited by the clock stability.

\begin{figure}[htbp]
    \centering
    \includegraphics[width=0.48\textwidth]{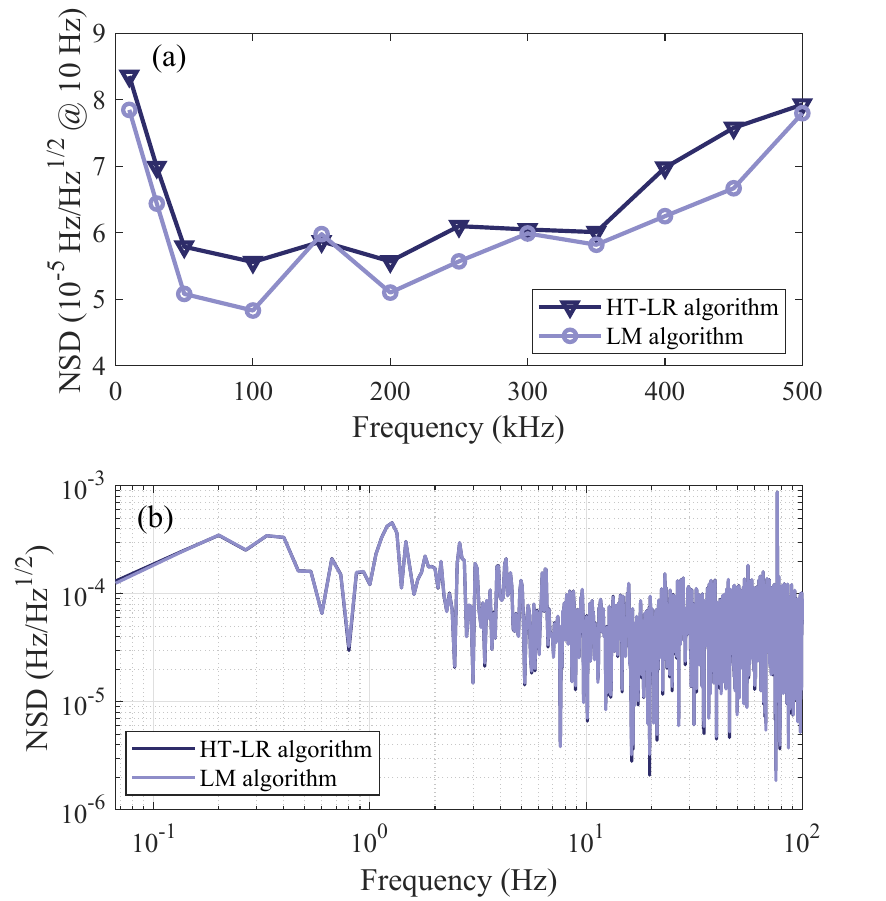}
    \caption{(a) Comparison between the HT-LR algorithm and Levenberg-Marquardt fitting at different signal frequencies, with $ K=20 $ in the HT-LR algorithm. (b) NSD spectrum curve of the signal (using 200 kHz as an example).}
   \label{fig4}
\end{figure}

Figure \ref{fig4} shows that the NSD value of real-time readout of oscillation frequencies of the pulsed input signals using our frequency counter is always better than 100 $\rm{\mu Hz/\sqrt {Hz}} $ at 10 Hz, which is very close to the off-line fitting results using the LM algorithm. It needs to be emphasized that the increase of NSD value at both low and high frequencies is not due to the limitations from the frequency counter or LM algorithm, but rather due to the reduced signal-to-noise ratio (SNR) at these frequencies of the specific signal generator (RIGOL DG2052) used in tests. We have tried to perform the same tests using a higher-precision signal generator (Zurich Instruments HDAWG), and within the frequency range of 10 kHz to 500 kHz, the NSD value of the extracted frequencies from the updated device is constantly between 10 $\rm{\mu Hz/\sqrt {Hz}} $ to 30 $\rm{\mu Hz/\sqrt {Hz}} $ at 10 Hz.

We have also tested the influence of the detection bandwidth on the device sensitivity. Although increasing the output rate leads to a reduced processing time available for each data segment, the number of sample points per segment of data also decreases. Therefore, with $ K = 20 $, the total computation time remains constant for different pulse repetition rates. 

\begin{figure}[htbp]
    \centering
    \includegraphics[width=0.48\textwidth]{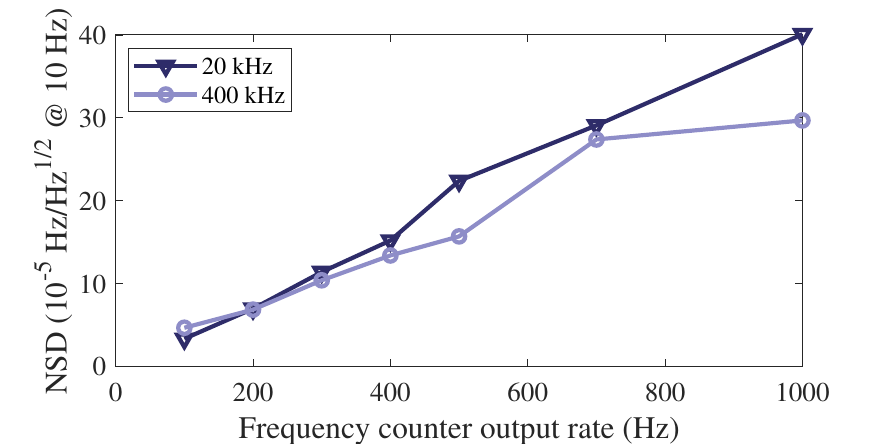}
    \caption{NSD of the frequency counter at different output rates.}
    \label{fig5}
\end{figure}

Figure \ref{fig5} shows that, as the output rate increases with all other parameters kept same in the tests, the amount of data involved in the computation decreases, leading to a corresponding decline in detection sensitivity. However, even at an output rate of 1000 Hz, the frequency sensitivity of our frequency counter remains below 400 $\rm{\mu Hz/\sqrt {Hz}} $ at 10 Hz.

\begin{figure}[htbp]
    \centering
    \includegraphics[width=0.48\textwidth]{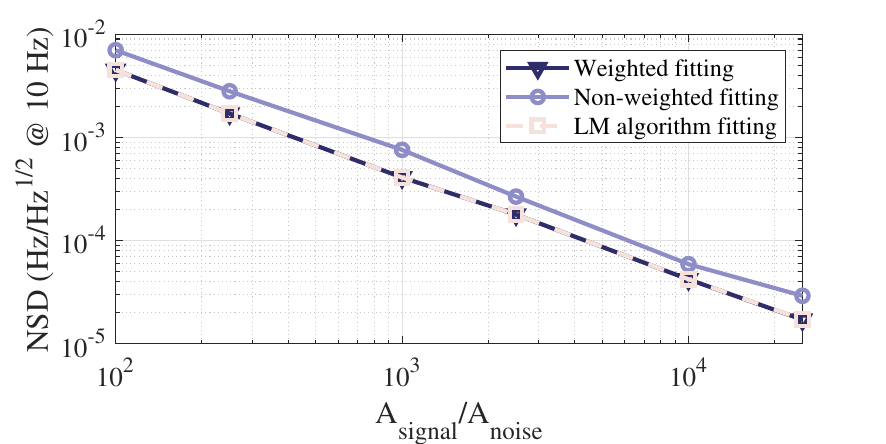}
    \caption{NSD of 250 kHz simulated signal processed by three algorithms (gate time $ T=5 $ ms).}
    \label{fig6}
\end{figure}

To increase the robustness of the HT-LR algorithm, we modify the linear fit part in the algorithm to a weighted fit, where the weights are determined by the amplitude of the analytic function $A(t)$ mentioned in Sec. II. Since the SNR of the input pulsed signal decreases with time, reducing the fitting weight of the later parts of the signal can improve the quality of the fitting. We test the performance of the weighted fitting procedure using simulated data. First, we generated a batch of 3000 data sets on the computer. The data takes the form

\begin{equation}
    \label{4}
    V[n]=A_{signal}e^{-\frac{n}{f_s\cdot \tau}}\sin (2\pi f\cdot n/f_s)+A_{noise}randn[n], 
\end{equation}

\noindent where $ \tau=2.5 $ ms, and $ f_s=1.53846 $ MSa/s. The frequency of the simulated data is set to 250 kHz, and $ randn[n] $ represents a set of standard normal distribution random numbers. We then process this data using both weighted and non-weighted HT-LR algorithms, as well as the LM fitting algorithms, comparisons of the resulted NSD are plotted in Fig. \ref{fig6}, and the result indicate that the weighted fitting procedure reduces the NSD by reducing the weight of low-SNR segments, with the improvement proportional to the gate time duration. Specifically, this improvement factor reaches approximately 15\% at $ T=2.5 $ ms, and escalates to 40\% when $ T $ is extended to 5 ms. Furthermore, the results from the weighted HT-LR algorithm are almost identical to those obtained using the LM fitting algorithm. Importantly, frequency deviations caused by SNR variations exhibit no significant correlation with SNR levels and are negligible compared to aforementioned truncation-induced deviations from parameter $ K $. This further validates the robustness of the weighted HT-LR algorithm in handling noise-contaminated signals.

According to theoretical analysis in Ref. \cite{Johansson2006}, for a particular frequency $f$, the fundamental sensitivity limit of TDC-based $\Omega$-counters is governed by

\begin{equation}
    \label{3.5}
    \delta f = \frac{2\sqrt{3} \cdot f^{1/2}}{T^{3/2}} \cdot \delta T,
\end{equation}

\noindent where \(\delta T\) denotes the TDC timing resolution. Assuming $\delta T = 20$ ps, a typical value in common commercial frequency counters, such as Keysight 53230A \cite{keysight}, and adopting our experimental gate time $T = 2.5$ ms, the $\Omega$-counter's theoretical sensitivity calculates to be 20 $\rm{\mu Hz/\sqrt {Hz}} $ with the signal oscillation frequency at 250 kHz, whereas our implementation achieves an NSD between 10 to 30 $\rm{\mu Hz/\sqrt {Hz}} $ across the entire frequency band (10-500 kHz) when using the high-precision Zurich Instruments HDAWG signal generator. Notably, this comparison assumes ideal conditions for the $\Omega$-counter: exhaustive timestamp recording for every signal cycle. We therefore conclude that the HT-LR scheme provides similar sensitivity to $\Omega$-counters within our target frequency band, while costing less hardware resources.

\section{Conclusion}

In conclusion, we have developed an algorithm based on Hilbert transform to extract the oscillation frequency of a pulsed signal, and also implemented this algorithm in a FPGA-based frequency counter. The real-time measurement results of this device demonstrates comparable sensitivity of extracted frequency with the commonly used off-line non-linear fitting method. With an output rate of 200 Hz, the measurement sensitivity is less than 100 $\rm{\mu Hz/\sqrt {Hz}} $ at 10 Hz, which corresponds to a background noise level of 10 $\rm{fT/\sqrt {Hz}} $ for $^{87}$Rb magnetometers. And even at an output rate of 1000 Hz, the measurement sensitivity is below 400 $\rm{\mu Hz/\sqrt {Hz}} $ at 10 Hz. Applications of such an algorithm and device are not limited in the field of atomic magnetometry, they can be applied to other scenarios requiring high-precision, real-time processing of pulse frequency signals. Following this work, we will focus on extending the algorithm to extract multiple frequencies and further optimizing the computational speed.

\section*{Acknowledgments}
This work was supported by National Natural Science Foundation of China (Grant No. 12174372), and Scientific Instrument and Equipment Development Projects, CAS (No. YJKYYQ20200043).


\end{document}